\documentclass[10pt,conference]{IEEEtran} 
\IEEEoverridecommandlockouts
 \makeatletter
\usepackage{cite}
\usepackage{amsmath,amssymb,amsfonts}
\usepackage{algorithmic}
\usepackage{graphicx}
\usepackage{textcomp}
\usepackage{xcolor}

\usepackage{tablefootnote}
\usepackage{subcaption}
\usepackage{graphicx}
\usepackage{float}

\usepackage{xcolor}
\usepackage{ifthen}

\usepackage{verbatim}

\usepackage{tikz}
\usepackage{collcell}

\newcommand*{\MinNumber}{0.0}%
\newcommand*{\MidNumber}{0.5} %
\newcommand*{\MaxNumber}{1.0}%

\newcommand{\ApplyGradient}[1]{%
             
        \ifdim #1 pt > \MidNumber pt
            \pgfmathsetmacro{\PercentColor}{max(min(100.0*(#1 - \MidNumber)/(\MaxNumber-\MidNumber),100.0),0.00)} %
            \hspace{-0.33em}\colorbox{green!\PercentColor!red}{#1}
        \else 

            \ifdim #1 pt > -0.1 pt
            \pgfmathsetmacro{\PercentColor}{max(min(100.0*(\MidNumber - #1)/(\MidNumber-\MinNumber),100.0),0.00)} %
            \hspace{-0.33em}\colorbox{red!\PercentColor!yellow}{#1}
             \else 
                \pgfmathsetmacro{\PercentColor}{max(min(100.0*(\MidNumber - #1)/(\MidNumber-\MinNumber),100.0),0.00)} %
                \hspace{-0.33em}\colorbox{white}{}

            \fi
        \fi

}

\newcommand{\ApplyGradientt}[1]{%
    \ifthenelse{#1 pt > \MidNumber pt}{\pgfmathsetmacro{\PercentColor}{max(min(100.0*(#1 - \MidNumber)/(\MaxNumber-\MidNumber),100.0),0.00)} %
            \hspace{-0.33em}\colorbox{green!\PercentColor!yellow}{#1} }{%
        \ifthenelse{#1 pt > 0 pt  }{\pgfmathsetmacro{\PercentColor}{max(min(100.0*(\MidNumber - #1)/(\MidNumber-\MinNumber),100.0),0.00)} %
            \hspace{-0.33em}\colorbox{green!\PercentColor!yellow}{#1}}{%
            \ifthenelse{#1 pt < 0 pt}{\pgfmathsetmacro{\PercentColor}{max(min(100.0*(\MidNumber - #1)/(\MidNumber-\MinNumber),100.0),0.00)} %
            \hspace{-0.33em}\colorbox{green!\PercentColor!yellow}{#1}}{%
    
            }%
        }%
    }%
}

\newcolumntype{R}{>{\collectcell\ApplyGradient}c<{\endcollectcell}}

\def\BibTeX{{\rm B\kern-.05em{\sc i\kern-.025em b}\kern-.08em
    T\kern-.1667em\lower.7ex\hbox{E}\kern-.125emX}}

\usepackage{authblk} %for \affil
\usepackage{mathtools}

\usepackage{stackengine}
\newcommand\underparen[1]{\@ifnextchar_{\uphelp{\uparen{#1}}}{\uparen{#1}}}
\makeatother
\def\uphelp#1_#2{\ensurestackMath{\stackunder[1pt]{#1}{\scriptstyle #2}}}
\newcommand\uparen[1]{\setbox0=\hbox{$#1$}\ensurestackMath{%
  \stackunder[0pt]{#1}{\rotatebox{90}{$\left(%
  \rule[\dimexpr-.5\wd0+\dp\strutbox-1.3pt]{0pt}{\wd0}\right.$}}%
}}

\usepackage{amssymb} % for triangleq

\usepackage{xspace}

\newcommand{\keepcomment}{1} % 1 - Keep comments, 0 - Hide comments

\usepackage[normalem]{ulem}

\ifnum\keepcomment=1
  \usepackage[colorinlistoftodos,textsize=scriptsize]{todonotes} % todo
    \newcommand{\stkout}[1]{\ifmmode\text{\sout{\ensuremath{#1}}}\else\sout{#1}\fi}
    
\else
  
  \usepackage[disable]{todonotes} % todo
\fi

\usepackage{paralist} % To use compactitem

\usepackage{diagbox}

\begin{document}

\title{Can Edge Computing fulfill the requirements of automated vehicular services using 5G
network ?

}

\author{Wendlasida Ouedraogo}
\author{Andrea Araldo}
\author{Badii Jouaber}
\author{Hind Castel}
\author{Remy Grunblatt}
\affil{Telecom SudParis, Institut Polytechnique de Paris; France; \{firstname.lastname\}@telecom-sudparis.eu}


\maketitle

\begin{abstract}

Communication and computation services supporting Connected and Automated Vehicles (CAVs) are characterized by stringent requirements, in terms of response time and reliability. Fulfilling these requirements is crucial for ensuring road safety and traffic optimization. The conceptually simple solution of hosting these services in the vehicles increases their cost (mainly due to the installation and maintenance of computation infrastructure) and may drain their battery excessively. Such disadvantages can be tackled via Multi-Access Edge Computing (MEC), consisting in deploying computation capability in network nodes deployed close to the devices (vehicles in this case), such as to satisfy the stringent CAV requirements. However, it is not yet clear under which conditions MEC can support CAV requirements and for which services. To shed light on this question, we conduct a simulation campaign using well-known open-source simulation tools, namely OMNeT++, Simu5G, Veins, INET, and SUMO. We are thus able to provide a reality check on MEC for CAV, pinpointing what are the computation capacities that must be installed in the MEC, to support the different services, and the amount of vehicles that a single MEC node can support. We find that such parameters must vary a lot, depending on the service considered. This study can serve as a preliminary basis for network operators to plan future deployment of MEC to support CAV.

\end{abstract}

\begin{IEEEkeywords}
 5G Simulation; MEC; Connected and Automated Vehicles
\end{IEEEkeywords}

\section{Introduction}
In order to safely and efficiently perceive their surroundings, make real-time decisions, and navigate complex environments, CAVs need some computational resources, e.g., CPU, GPU, and RAM, to perform some algorithmic tasks. Computational requirements may fluctuate depending on road traffic conditions. Installing a large amount of computational resources to be ready to perform peak computation at any moment is very costly and would make vehicles too expensive, thus severely restricting the market for CAV. Moreover, a large amount of computation consumes power, thus reducing vehicle autonomy or requiring big, heavy, and expensive batteries. Cloud Computing partially addresses this problem by offering offloading capabilities. However, CAVs have stringent requirements regarding latency and bandwidth that can hardly be met by Cloud Computing\cite{CLOUDDELAY}. Hence, MEC emerges as an alternative capable of meeting those requirements, thanks to deploying computational resources very close to vehicles, i.e., in the base stations or in the Road Side Units (RSUs). Also, the reliability, the high bandwidth, and the extensive coverage offered by 5G networks, together with the integration of MEC, offer promising conditions for a large penetration of CAVs. In this context, a question remains open: 

\emph{With which computational capacity should MEC nodes be equipped in order to support the different CAV services?} 

To answer this question, we first present a basic queueing theory model to find the lower bounds on the required computational resources (\S\ref{sec:performance}). We then perform a simulation campaign (\S\ref{sec:simulation-environment}-\ref{sec:results}). We simulate the realistic mobility of vehicles using SUMO. Packet exchanges between vehicles and the MEC, as well as CAV service computation, are simulated using essentially  OMNeT++\cite{OmnetGithub}, Simu5G\cite{Simu5gGithub}, and Veins\cite{VeinsGithub}. Our code, released in open source\footnote{https://github.com/zazim13/Simu5G-MecBasedAV} can be used by other researchers as a base to study the feasibility of MEC-based CAV services.

We find that for some CAV services, such as remote driving and cooperative sensing, the amount of required MEC resources is high, and only a few vehicles can be supported at the same time by a MEC node, making MEC a difficult avenue to follow. For other CAV services such as cooperative maneuver and awareness, instead, MEC is more promising, being able to support a larger number of vehicles.

\section{Background and motivation}

CAVs are categorized into different levels of automation, as defined by the Society of Automotive Engineers (SAE)\cite{LevelOfAutomation}. These levels range from level 0 (no automation) to level 5 (full automation). For this work, we consider scenarios with a higher level of automation (from Level~3 to~5), which means that vehicles can perform the majority of driving tasks, such as platooning, collision avoidance, or cooperative sensing. 

In these scenarios, vehicles communicate using vehicle-to-everything (V2X) communications, which can be with other vehicles (vehicle-to-vehicle or V2V), the infrastructure (vehicle-to-infrastructure or V2I), pedestrians (vehicle-to-pedestrians or V2P), or the network (vehicle-to-network or V2N). Performing in the edge node computation pertaining to the aforementioned driving tasks relies on V2N communications, enabling communication with the MEC. In this case, we talk about vehicle-to-edge and edge-to-vehicle communications.

Moving a large part of driving-related computation from vehicles to edge nodes provides several advantages. Such advantages have been shown for platooning~\cite{MANCUSO}, but they extend, more generally, to various CAV applications. Notably, the MEC facilitates interoperability among diverse CAV hardware, enabling a central controller to simplify collaboration between different vehicle types. In addition, edge nodes can be made more resilient by appropriate redundancy practices. If appropriately positioned, edge nodes can be less susceptible to shadowing when communicating with vehicles. Finally, an edge node can aggregate information from multiple vehicles, fostering enhanced cooperation and facilitating global decision-making, as showcased in~\cite{Edge_powered} and~\cite{VECFrame}.

However, while the MEC provides improved latency compared to the Cloud, it does introduce some latency in contrast to on-board computations. Studies such as \cite{vehicle_control} suggest balancing the computational load between the edge node and the Cloud, based on latency considerations. In our approach, we only focus on leveraging the MEC, as our goal is to comprehend its limitations in supporting CAV services.

\section{Vehicular services and architecture}
 
\subsection{Vehicular applications}
\label{section:use_cases_definition}

%aa: The sentence below is useless

%In this section we clearly state services or use cases that we consider for our simulations. 

%aa: The sentence below is useless
%A variety of use cases exists as there are a multitude of scenarios to address. In this paper, w

We consider the following CAV services, related to the high degree of automation, characterized by stringent requirements, as described in \cite{CONNECTEDROADS}.
\begin{itemize}

    \item \emph{Remote driving} allows vehicles to be driven by a human operator or an application outside the vehicle. 
    \item \emph{Cooperative sensing} involves exchanging sensed data to enhance a vehicle's environmental perception.
    \item \emph{Cooperative maneuver} consists of exchanging messages to synchronize vehicles' maneuvers, such as lane changing or platooning.
    \item \emph{Cooperative awareness}: information exchange to inform vehicles about relevant events, such as Emergency Vehicle Alerts, electronic emergency brake signals, etc.
\end{itemize}

For these use cases, the acceptable requirements are provided in \cite{CONNECTEDROADS} and ~\cite{RequirementDefinition}. Table~\ref{tab:Use_cases_requirements} summarizes these requirements. The requirements pertain to \textbf{end-to-end latency}, signifying the delay from the transmission of a message to its reception at the application level, and \textbf{reliability}, indicating the probability of successfully transmitting data within the end-to-end delay threshold, as in~\cite{RequirementDefinition}.

\begin{table}[]
\caption{Use cases requirements}  
\label{tab:Use_cases_requirements}
\centering
\small
\setlength\tabcolsep{2pt}
\resizebox{0.9\columnwidth}{!}{
 
    \begin{tabular}{|c|c|c|} 
        \hline 
         Use cases&  End-to-End Latency Threshold&  Reliability \\ 
         \hline 
         Remote driving&  20ms& 99\%\\ 
         \hline 
         Cooperative sensing&   10ms&  95\%\\ 
         \hline 
         Cooperative maneuver& 100ms&  99\%\\ 
         \hline 
         Cooperative awareness&  100ms& 95\%\\ 
         \hline
    \end{tabular}   
    
    }
    
\end{table}

\subsection{Communication behavior}
  
Our objective is to assess the feasibility of deploying applications in the MEC to serve the use cases described in \S\ref{section:use_cases_definition}. Since our paper is not a study on the internals of the aforementioned applications, we made a deliberate decision not to implement each specific application in detail, which would be time-consuming. Instead, we simulate a generic parametric application. By adjusting the parameters to match the application characteristics (as in Table~\ref{tab:use_cases_specifications}), we can mimic the behavior of the different applications at a high level, only focusing on network-related characteristics. By mimicking, we mean generating communications that such an application would perform, consuming the same bandwidth, with the same data rate, and demanding the same processing load. We consider two communication behaviors that represent the most distinct services.
 
\begin{figure}[]
\caption{Communication behavior of type ``Dissemination''.}
 \includegraphics[width=2.3in]{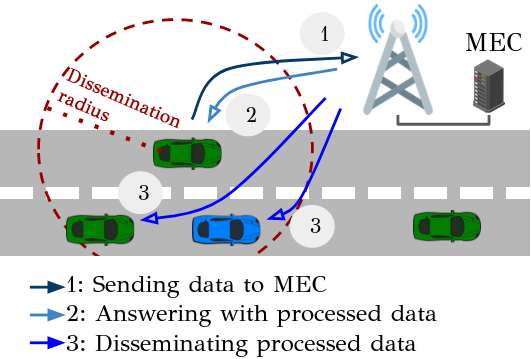}
\label{fig:dissemination}
\centering
\end{figure}

\textbf{\emph{Dissemination}}. For cooperative applications, data such as cooperative sensing, warning alerts, or traffic flow information needs to be shared with nearby vehicles or RSUs. Our approach emphasizes vehicle-to-edge and edge-to-vehicle communication, following the methodology in \cite{MANCUSO} for platooning services. In this model, the edge node processes data, runs algorithms, and disseminates results to relevant vehicles. We introduce the 'dissemination radius' parameter, creating a circle around the data-producing vehicle for information sharing. Each vehicle wishing to cooperate sends its data to the edge node, which processes it and shares the information in a unicast manner with all vehicles within the dissemination circle. While broadcast is an option, it complicates limiting dissemination to the specified circle. This behavior is depicted in Fig.~\ref{fig:dissemination}.

\textbf{\emph{Client-Server}}. For remote driving service, the edge node is generally used to offload certain tasks, such as object recognition and parking assistance (as in~\cite{MIfog}). In this paper, we assume that the edge node is completely in charge of remotely operating cars. This means that remotely driven vehicles send data (videos, LiDAR, or sensed data) to the edge node, which handles the execution of the necessary tasks and sends the right commands, e.g., steering, braking, and accelerating.

 We choose a proactive behaviour for all applications, meaning that data are periodically sent to the edge node then shared when needed, and deploying the service is trigering the sending of data, no particular event is waited.

\begin{comment}

\subsection{MEC architecture}
\label{sec:MEC-architecture}

Following~\cite{Rapid_prototyping}, a service used by a vehicle consists of a UEApp and a MecApp. The former is a virtualized application (e.g., running in a container) running within the vehicle, typically in its On-Board Unit (OBU). The MecApp is a virtualized application running in the edge node. Observe that there is a MecApp running in the edge node for each service and each vehicle. In other words, each vehicle service has a corresponding counterpart running in the edge node.
Therefore, when a vehicle wants to start to use a service, it launches the corresponding UEApp and requests the instantiation of the corresponding MecApp in the edge node. Each MecApp has a minimum CPU requirement. If enough resources are available in the edge node, the MecApp can be instantiated and the service can successfully run, based on the communication between UEApp and MecApp. All aspects of this communication, such as data rate, payload size, etc., are parameters used to simulate actual CAV scenarios, according to Table~\ref{tab:use_cases_specifications}. If not enough resources are available in the edge node, the vehicle cannot run the service.

\end{comment}

%%%%%%%%%%%%%%%%%%%%%%%%%%%%%%%
%%%%%%% PERFORMANCE %%%%%%%%%%%
%%%%%%% CHARACTERIZATION %%%%%%
\section{Performance characterization}
\label{sec:performance}
\begin{figure}[]
\caption{Delay components}
 \includegraphics[width=3in]{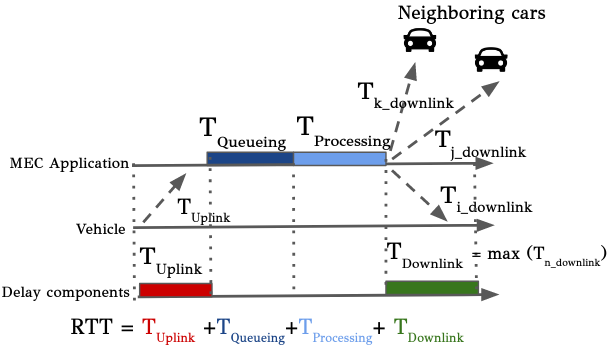}
\label{fig:delay_component}
\centering
\end{figure}

The total delay is composed of several components, as depicted in Fig.~\ref{fig:delay_component}. Each application running in the MEC (MecApp), related to one service and one vehicle, is allocated a limited amount of processing resources. For the sake of simplicity, we here only focus on the CPU deployed in an edge node. We assume that the CPU of the edge node is allocated among several MecApps using a fair sharing discipline~\cite{Rapid_prototyping}. A MecApp operates as a queueing system for incoming packets. Queued packets are processed sequentially, with the processing time determined by the allocated CPU and task complexity. The CPU allocation directly influences both processing time and end-to-end delay, a critical factor for meeting reliability requirements in the context of automated vehicles.
% \begin{figure}[]
%\caption{MecApp modeled as a queueing system}
% \includegraphics[width=\linewidth]{figures/queue.png}
%\label{fig:mecAppQueue}
%\centering
%\end{figure}

If we aim to achieve a reliability of \(R_{req}\), meaning that more than \(R_{req}\) percent of the packets should be successfully transmitted within the end-to-end delay requirement \(D_{req}\), we must ensure that for an end-to-end delay \(D\).
\begin{equation}
P(D \leq D_{req}) \ge R_{req}
\label{eqD}
\end{equation}

 From \S\ref{specifications} the packet generation follows a Poisson process, and we also assume that the arrival rate at the MecApp follows a Poisson distribution. In addition, the processing time is assimilated to an exponential distribution. Given that,  each MecApp can be assimilated to a queueing system, we can approximate these queues as M/M/1 queues. Consequently, the time taken by each packet within the MecApp (queueing and processing delay) follows an exponential distribution, with a parameter \(\mu-\lambda\), where \(\lambda\) represents the packet arrival rate at the MecApp and \(\mu\), the processing rate (in packets/s).

To ensure that the requirement can be met, it's crucial that the time taken by each packet within the MecApp \(D_{mec}\) respects the condition (\ref{eqD}) i.e. a necessary condition to meeting the reliability requirement is : 
\begin{equation}
P(D_{mec} \leq D_{req})\geq R_{req}
\label{eqDmec}
\end{equation}

then we obtain:

\begin{equation}
\mu \ge \lambda - \frac{ln(1-R_{req})}{D_{req}}
\end{equation}
 where \(R_{req}\) is the overall reliability requirement and \(D_{req}\) is the overall delay requirement.

 Given that: 
 
 %\begin{equation}
%\mu = \frac{E(IPR)}{Allocated_{CPU}}     
% \end{equation} 

%\begin{equation}
%\mu = \frac{Allocated_{CPU}}{E(IPR) }    
 %\end{equation} 
\begin{equation}
\mu = \frac{CPU}{E(IPR) }    
 \end{equation}

Where $CPU$ is the amount of CPU (in instructions/s) allocated to the MecApp and \(IPR\) is defined in \S\ref{ipr}. The minimum CPU that should be allocated to each MecApp in order to respect the necessary condition (\ref{eqDmec}) is:

\begin{equation}
CPU_{min} = \left( \lambda - \frac{\ln(1-R_{req})}{D_{req}} \right) \cdot E(IPR)
\end{equation}

We compute these minimum CPU allocations regarding the requirements presented in Table~\ref{tab:Use_cases_requirements}. They are summarized in Table~\ref{tab:use_cases_specifications}. Observe a notable discrepancy among the CPU required for different services.

\begin{comment}

\section{Simulation implementation}
\label{sec:simulation-implementation}

Using OpenStreetMap and SUMO tools, we generate a real-world road map corresponding to a district of Paris of 1 square-kilometer, as depicted in Fig.~\ref{fig:architecture}. We also generate traffic demand, and each vehicle is a module whose mobility is defined by SUMO. Each vehicle is equipped with a network interface card that implements the 5G New Radio stack, allowing the vehicle to connect to a gNB. In our infrastructure, we place a central gNB, a MEC system (a UALCMP, a MEC orchestrator, and a MEC host) compliant with the ETSI model\cite{etsimec}. Modules are connected through a UPF and an intermediary UPF. The wired connections are 10 Gigabit/sec Ethernet links, provided by INET.

Simu5G natively models the MEC architecture of \S\ref{sec:MEC-architecture}, as described in~\cite{Rapid_prototyping}. All the code is released as open source in \cite{MecBasedAV}, a forked repository of Simu5G with a new scenario featuring the described application.

\end{comment}

\section{Simulation environment}
\label{sec:simulation-environment}
\begin{figure}[]
\caption{Simulation frameworks diagram}
 \includegraphics[width=2in]{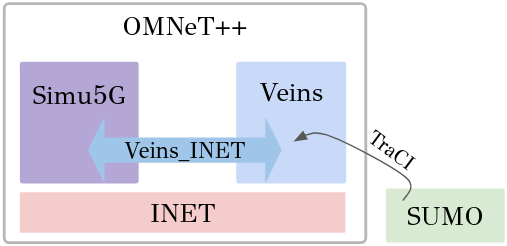}
\label{fig:Frameworks}
\centering
\end{figure}

We use several open-source software and frameworks to perform our simulations. Fig.~\ref{fig:Frameworks} illustrates the diagram of interconnectivity within these frameworks. Our work is mainly based on \textit{Simu5G} \cite{Simu5gGithub}, a popular open-source 5G simulation library, because it provides 3GPP-compliant 5G New Radio Access and integrates MEC features. \textit{Simu5G} is based on the \textit{OMNeT++} framework \cite{OmnetGithub} and integrates \textit{INET} \cite{InetGithub}, which is an open-source framework providing tools for communication network simulation. In order to enable a vehicular network with realistic mobility, we use \textit{SUMO}\cite{SUMO2018}. \textit{SUMO} allows us to generate road networks and traffic demand, specifying routes and vehicles with realistic behavior. \textit{SUMO} provides an API \cite{traci} for interacting with the vehicles it generates, and \textit{Veins} (especially its subproject \textit{Veins\_INET}), another open-source vehicular network simulation framework allows us to access this API within the \textit{OMNeT++} simulation environment and give to vehicles in Simu5G the realistic mobility provided by \textit{SUMO}. With these tools, we can simulate scenarios aligning with our vehicle-to-edge and edge-to-vehicle communication approach.

\subsection{Simulation parameters}

\subsubsection{Radio conditions}
Simu5G offers a number of parameters to tune the 5G New Radio network. It also provides a realistic channel model that supports fading, path loss, shadowing, and attenuations. We choose to use macro base stations in an urban scenario. Other parameters are summarized in Table~\ref{tab:radio_parameter} and are essentially taken from \cite{radioParameters}.

\begin{table}[]
  \caption{Simulation radio parameters}
  \label{tab:radio_parameter}
  \centering
  \small
        \setlength\tabcolsep{1pt}
  \resizebox{0.7\columnwidth}{!}{\begin{tabular}{|l|l|}
  \hline
    Parameter name & Value \\   \hline
    Number of gNBs   & 1\\  \hline
    Carrier frequency & 6GHz \\  \hline
    Bandwidth & 80MHz(100 PRBs)\\ \hline 
    Numerology&2\\\hline  
    Fading (Jakes) + shadowing & enabled \\  \hline
    gNB Tx power & 46 dBm \\  \hline
    gNB antenna gain & 8dBi \\  \hline
    gNB noise figure &  5dB  \\  \hline
    UE antenna gain & 0dBi \\  \hline
    UE noise figure & 7dB \\  \hline
    Path loss model & (3GPP - TR 36.873) \\  \hline
 Blershift&5\\\hline
   \end{tabular}}
\end{table}

\subsubsection{Use cases specifications}
\label{specifications}
Our generic application aims to mimic real-world application behaviors. To achieve this objective, we studied these behaviors in the literature and collected specifications such as distributions, values, and other metrics for parameters like uplink or downlink bandwidth usage, data rates, and more. These specifications are summarized in Table~\ref{tab:use_cases_specifications}.

It is generally assumed in the literature, as in \cite{IoT_dist}, that the packet generation process follows a Poisson distribution for IoT-based applications. Then, assuming a constant data accumulation by IoT devices, the payload size of each sent packet can be approximated as an exponential random distribution. 

The processing time is directly related to the size of the packet to be processed, which follows an exponential distribution, we then also assume that the processing time itself is exponentially distributed. In Simu5G's MEC implementation, it is through instructions required by the packet being processed that the processing time is tuned, as they are proportional. For each use case, we consider that each packet requires certain tasks to be performed. We introduce the parameter \(IPR\)\label{ipr} for Instructions Per Request i.e. the number of instructions required per each request. We detailed in Table~\ref{tab:use_cases_specifications} values of \(IPR\) for each use case. We extract them from \cite{MIfog}, where we can find task instructions requirements for tasks related to automated vehicles. Each use case is attributed with sufficient \(IPR\) to handle essential tasks such as steering control, alongside characteristic tasks corresponding to each use case (e.g., object recognition for remote driving or parking assistance for cooperative maneuvers). We use MIPS (Millions of Instructions Per Second) and MI(Million Instructions) as a measure of computational capability and number of instructions to align with Simu5G.

Once processed, an answer with the processed information is transmitted back to the sending vehicle, and in the case of cooperative applications, to neighboring vehicles as well. These transmitted packets include processed information, such as steering control, alerts, acknowledgments, and more. In \cite{AUTOMATEDVECH}, a downlink bandwidth of 0.25 MBit/s is assumed for self-driven cars, with downlink packets containing steering control information. With a sending rate of 100 messages per second in our remote driving scenarios, this implies an answering rate of 100 messages per second. Consequently, we assume a payload size of 313 bytes for each packet, considering it to be adequate to accommodate all potential types of responses across various use cases.

Each use case is defined by a communication range, representing the area over which V2X messages can be transmitted. We adopt a circle of dissemination approach, where the radius of this circle determines the communication perimeter. We assume that this radius follows a uniform distribution within the specified range detailed in \cite{CONNECTEDROADS}, as indicated in Table~\ref{tab:use_cases_specifications}.

\begin{table*}
    \centering
      \small
        \setlength\tabcolsep{2pt}
\caption{Use cases specifications}  
\label{tab:use_cases_specifications}
    \resizebox{\textwidth}{!}{\begin{tabular}{|l|l|l|l|l|l|l|c|} \hline 
         Use cases& Uplink bandwidth
(Mb/s)&Uplink rate (msg/s)&  Uplink payload (Bytes)&  Downlink payload (Bytes)&IPR \cite{MIfog}&Dissemination radius& Min CPU(MIPS) \\ \hline 
         Remote driving& 32 
\cite{CARLA}&Pois(100)&  Exp(40000)&  313  \cite{AUTOMATEDVECH}&Exp(500)&-&165130\\ \hline 
         Cooperative sensing& 10  \cite{CONNECTEDROADS}&Pois(100)  \cite{ITSCAM}&  Exp(12500)&  313  \cite{AUTOMATEDVECH}&Exp(200)&Uniform(0-200m) \cite{CONNECTEDROADS}&79915\\ \hline 
         Cooperative maneuver& 1.3 \cite{CONNECTEDROADS}&Pois(10) \cite{MANCUSO}&  Exp(16250) &  313  \cite{AUTOMATEDVECH}&Exp(500)&Uniform(0-500m) \cite{CONNECTEDROADS}&28026\\ \hline 
         Cooperative awareness& 0.12 &Pois(10) \cite{ITSCAM}&  Exp(1500) \cite{CONNECTEDROADS}&  313  \cite{AUTOMATEDVECH}&Exp(200)&Uniform(0-500m) \cite{CONNECTEDROADS}&7992\\ \hline
    \end{tabular}}
\end{table*}

%\subsubsection{MEC and resources allocation}
%\label{queue}

\subsubsection{Simulation scenario}
In our simulations, we essentially tune two parameters: the CPU capacity available at the edge, shared among the MecApps of the cars, and the number of injected vehicles. The processor used and their computational capacities are displayed in Tables~\ref{table:processing_power}. Each experiment corresponds to a simulation involving a specific edge node processor and a certain number of injected vehicles. Every experiment is repeated five times with different seeds (i.e., five repetitions in OMNeT++), and each simulation lasts 180 seconds. The road map corresponds to a district of Paris of 1 square kilometer.

\begin{table}[]
        \caption{Processor Processing Speed}
        \label{table:processing_power}
        \begin{tabular}{|l|l|l|c|}
            \hline
            \multicolumn{2}{|c|}{Id}&Processor& Processing Speed(MIPS)\cite{wikipedia_MIPS}\\ \hline
            \multicolumn{2}{|l|}{1}&AMD Ryzen Threadripper&2356230 \\ \hline
            \multicolumn{2}{|l|}{2}&AMD Ryzen 9&749070\\ \hline
            \multicolumn{2}{|l|}{3}&Intel Core i9-9900K&412090  \\ \hline
            \multicolumn{2}{|l|}{4}&Intel Core i5-11600K&346350 \\ \hline
        \end{tabular}
\end{table}

\begin{comment}    

\begin{table}
    \centering
     \small
        \setlength\tabcolsep{2pt}
\caption{Simulation parameters}    
\label{table:Simulation_parameters}
    \resizebox{0.7\columnwidth}{!}{\begin{tabular}{|c|c|} \hline 
         Use cases&  Vehicles injected\\ \hline 
         Remote driving&  1,2,3,4,5,6,7\\ \hline 
         Cooperative sensing&  1,2,3,4,5,6,7\\ \hline 
         Cooperative maneuver&  1,10,20,30,40,50,60\\ \hline 
         Cooperative awareness&  1,20,40,60,80,100,120\\ \hline
    \end{tabular}  }
\end{table}
\end{comment}

%%%%%%%%%%%%%%%%%%%%%%%%%%%
%%%%%%%%% RESULTS %%%%%%%%%
%%%%%%%%%%%%%%%%%%%%%%%%%%%
\section{Results}
\label{sec:results}

 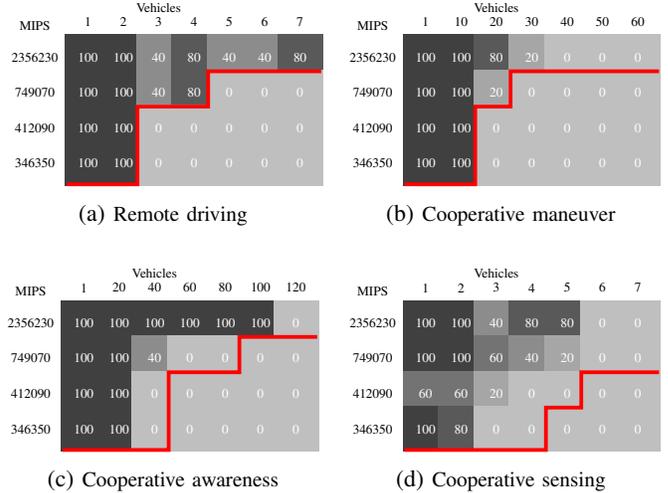
\begin{figure}
 
        \centering
        \begin{subfigure}[]{0.493\linewidth}
            \centering     
                        \begin{tikzpicture}[scale=0.47]
                        \node[above,font=\tiny] at (3,0) {Vehicles};
                        \node[above,font=\tiny] at (-0.5,-0.5) {MIPS};
                        
                          \foreach \y [count=\n] in {
                              {100,100,40,80,40,40,80},
                              {100,100,40,80,0,0,0},
                              {100,100,0,0,0,0,0},
                              {100,100,0,0,0,0,0},
                            } {
                        
                              % heatmap tiles
                              \foreach \x [count=\m] in \y {
                                \node[fill=darkgray!\x!lightgray, minimum size=6mm, text=white] at (\m,-\n) {\tiny \x};
                              }
                            }
                          % % column labels
                          \foreach \a [count=\i] in {1,2,3,4,5,6,7} {
                            \node[minimum size=6mm] at (\i,0) {\tiny \a};
                          }
                          % row labels
                          \foreach \a [count=\i] in {2356230,749070,412090,346350}{
                            \node[minimum size=6mm] at (-0.5,-\i) {\tiny \a};
                          }

    \draw[line width=0.5mm,red] (0.4,-4.6) -- (1.4,-4.6) -- (2.4,-4.6) -- (2.4,-2.4) -- (4.4,-2.4) -- (4.4,-1.4) -- (7.6,-1.4);
    
   % \draw[line width=0.5mm,blue]  (6.4,-1.3)--(6.4,-0.4) -- (4.4,-0.4) -- (4.4,-1.3) -- (6.4,-1.3);

    %\draw[line width=0.5mm,blue]  (2.4,-2.3) -- (3.4,-2.3)--(3.4,-0.4) -- (3.4,-0.4) -- (2.4,-0.4) -- (2.4,-2.3);
                        \end{tikzpicture}
            \caption[]%
            {{\footnotesize Remote driving}}    
        \end{subfigure}
        \hfill
        \begin{subfigure}[]{0.493\linewidth}  
            \centering 
                        \begin{tikzpicture}[scale=0.47]
                         \node[above,font=\tiny] at (3,0) {Vehicles};
                        \node[above,font=\tiny] at (-0.5,-0.5) {MIPS};
                          \foreach \y [count=\n] in {
                              {100,100,80,20,0,0,0},
                              {100,100,20,0,0,0,0},
                              {100,100,0,0,0,0,0},
                              {100,100,0,0,0,0,0},
                            } {
                        
                              % heatmap tiles
                              \foreach \x [count=\m] in \y {
                                \node[fill=darkgray!\x!lightgray, minimum size=6mm, text=white] at (\m,-\n) {\tiny \x};
                              }
                            }
                          % % column labels
                          \foreach \a [count=\i] in {1,10,20,30,40,50,60} {
                            \node[minimum size=6mm] at (\i,0) {\tiny \a};
                          }
                          % row labels
                          \foreach \a [count=\i] in {2356230,749070,412090,346350}{
                            \node[minimum size=6mm] at (-0.5,-\i) {\tiny \a};
                          }
                           \draw[line width=0.5mm,red] (0.4,-4.6) -- 
                           (1.4,-4.6) -- (2.4,-4.6) -- (2.4,-2.4) -- (3.4,-2.4) -- (3.4,-1.4) -- (7.6,-1.4) ;

                           %\draw[line width=0.5mm,blue]  (7.6,-1.3) -- (7.6,-0.4) -- (3.4,-0.4) -- (3.4,-1.3) -- (7.6,-1.3) ;

                           %\draw[line width=0.5mm,blue]  (3.3,-1.4) -- (2.4,-1.4) -- (2.4,-2.3) -- (3.3,-2.3) -- (3.3,-1.4) ;
                        \end{tikzpicture}
            \caption[]%
            {{\footnotesize Cooperative maneuver}}    
        \end{subfigure}
        \vskip\baselineskip
        \begin{subfigure}[]{0.493\linewidth}   

                        \begin{tikzpicture}[scale=0.47]
                         \node[above,font=\tiny] at (3,0) {Vehicles};
                        \node[above,font=\tiny] at (-0.5,-0.5) {MIPS};
                          \foreach \y [count=\n] in {
                              {100,100,100,100,100,100,0},
                              {100,100,40,0,0,0,0},
                              {100,100,0,0,0,0,0},
                              {100,100,0,0,0,0,0},
                            } {
                        
                              % heatmap tiles
                              \foreach \x [count=\m] in \y {
                                \node[fill=darkgray!\x!lightgray, minimum size=6mm, text=white] at (\m,-\n) {\tiny \x};
                              }
                            }
                          % % column labels
                          \foreach \a [count=\i] in {1,20,40,60,80,100,120} {
                            \node[minimum size=6mm] at (\i,0) {\tiny \a};
                          }
                          % row labels
                          \foreach \a [count=\i] in {2356230,749070,412090,346350} {
                            \node[minimum size=6mm] at (-0.5,-\i) {\tiny \a};
                          }
                           \draw[line width=0.5mm,red] (0.4,-4.6) -- (2.4,-4.6) -- (3.4,-4.6) -- (3.4,-2.4) -- (5.4,-2.4) -- (5.4,-1.4) -- (7.6,-1.4); 

                           %\draw[line width=0.5mm,blue](7.6,-1.3) -- (7.6,-0.4) -- (6.4,-0.4) -- (6.4,-1.3) -- (7.6,-1.3);

                           %\draw[line width=0.5mm,blue] (5.3,-1.4) -- (2.4,-1.4) -- (2.4,-4.5) -- (3.3,-4.5) -- (3.3,-2.3) -- (5.3,-2.3) -- (5.3,-1.4);
                        \end{tikzpicture}
 
            \centering 
            \caption[]%
            {{\footnotesize Cooperative awareness}}    
        \end{subfigure}
        \hfill
        \begin{subfigure}[]{0.493\linewidth}   
            \centering 
                        \begin{tikzpicture}[scale=0.47]

                         \node[above,font=\tiny] at (3,0) {Vehicles};
                        \node[above,font=\tiny] at (-0.5,-0.5) {MIPS};
                          \foreach \y [count=\n] in {
                              {100,100,40,80,80,0,0},
                              {100,100,60,40,20,0,0},
                              {60,60,20,0,0,0,0},
                              {100,80,0,0,0,0,0},
                            } {
                        
                              % heatmap tiles
                              \foreach \x [count=\m] in \y {
                                \node[fill=darkgray!\x!lightgray, minimum size=6mm, text=white] at (\m,-\n) {\tiny \x};
                              }
                            }
                          % % column labels
                          \foreach \a [count=\i] in {1,2,3,4,5,6,7} {
                            \node[minimum size=6mm] at (\i,0) {\tiny \a};
                          }
                          % row labels
                          \foreach \a [count=\i] in {2356230,749070,412090,346350}{
                            \node[minimum size=5mm] at (-0.5,-\i) {\tiny \a};
                          }
                           \draw[line width=0.5mm,red] (0.4,-4.6) -- (4.4,-4.6) -- (4.4,-3.4) -- 
                           (5.4,-3.4) -- 
                           (5.4,-2.4) -- (7.6,-2.4); 
                           
                           %\draw[line width=0.5mm,blue] (2.4,-4.5) -- (4.3,-4.5)-- (4.3,-3.3) -- (5.3,-3.3) -- (5.3,-2.3) -- (7.6,-2.3) -- (7.6,-0.4) -- (5.4,-0.4) -- (5.4,-1.4) -- (3.4,-1.4) -- (3.4,-2.4) -- (2.4,-2.4) -- (2.4,-4.5); 
                           
                           %\draw[line width=0.5mm,blue] (3.4,-1.4) -- (3.4,-0.4) -- (2.4,-0.4) -- (2.4,-1.4) -- (3.4,-1.4) ; 

                        \end{tikzpicture}
            \caption[]%
            {{\footnotesize Cooperative sensing}}    
        \end{subfigure}
        \caption []
        {\small Success rate of each Edge CPU } 
        \label{fig:heatmap}
    \end{figure}
Figure~\ref{fig:heatmap} illustrates the success rates across various experiments, where the success rate is defined as the percentage of repetitions meeting the reliability requirement out of all experiment repetitions. The red limitation denotes a threshold where CPU allocation meets the minimum requirements for each MECApp as outlined in Table~\ref{tab:use_cases_specifications}. Simulations below this threshold fail to meet the reliability criteria. %thereby validating our criterion as a necessary condition for success. 
However, simulations above this threshold show improved performance but still encounter challenges despite sufficient CPU allocation for each MECApp. This issue stems not only from delays inherent to the MEC node performance, addressed by the condition in (\ref{eqDmec}) which specifies the minimum CPU allocation required for each MEC application, but also from additional uplink and downlink delays. An increase in the number of vehicles leads to reduced bandwidth per vehicle, consequently increasing the end-to-end delay and highlighting the need for scaling edge nodes in proportion to network resources.

With their speed, processors id3 and id4 can effectively handle scenarios such as remote driving or cooperative sensing for up to 2 vehicles, awareness tasks for 20 vehicles, and maneuvering tasks for 10 vehicles. However, even doubling the processing speed with processor id2 does not significantly increase the number of vehicles that can be managed, with only a modest improvement observed in remote driving and cooperative sensing services. This underscores the challenge of scaling up the number of vehicles through improving the MEC node performance.

Moreover, our observations indicate that services with more stringent requirements support fewer vehicles given the same edge node capacity. Specifically, services like remote driving and cooperative sensing, which demand lower end-to-end delays, can support fewer vehicles than cooperative maneuver and awareness service. The latter services are therefore more promising.

%In our simulations, we examined various processors from ... to ...  
%based on parameters chosen in \cite{edge_400}, a processing capability of 400,000 MIPS appears to be acceptable for edge node processors. Processors with identifiers id3 and id4 are therefore deemed suitable for edge nodes or at least represent a reserved capacity for vehicular services.

\section{Conclusion}

We conducted a simulation campaign using diverse tools to assess the limits of integrating 5G networks and edge computing via the MEC paradigm to handle various CAVs services requirements. Our focus includes evaluating the feasibility of edge-based control, where MEC manages all required computations. We show a basic queuing model to provision the minimum computational capability to satisfy different MEC application requirements. Our results highlight a discrepancy in required resources for specific CAV services. This indicates that even if controlling CAVs from the edge is feasible, the efficiency greatly varies depending on the service. Additionally, we reveal challenges in improving such efficiency by scaling up the edge, as the communication delay can hinder the correct operation of vehicular services from the edge. A limitation of our approach is that we did not conduct an in-depth study of the communication delay nor did we consider cloud or vehicle computational resources. We plan to incorporate these factors in our future works to provide a more comprehensive analysis.

\section{Acknowledgments}
This work was carried out in the context of Beyond5G, a project funded by the French government as part of the economic recovery plan, namely "France Relance", and the investments for the future program.
\bibliographystyle{IEEEtran}
\bibliography{main}

\end{document}